\documentclass[letterpaper, 10 pt, conference]{ieeeconf}
\IEEEoverridecommandlockouts
\usepackage{graphics} % for pdf, bitmapped graphics files
\usepackage{amsmath} % assumes amsmath package installed
\usepackage{amssymb}  % assumes amsmath package installed
\usepackage{graphicx}
\usepackage{booktabs}
\usepackage{enumerate}

\usepackage{url}

\newcommand{\ubar}[1]{\mkern3mu\underline{\mkern-3mu #1\mkern-3mu}\mkern3mu}
\newcommand{\obar}[1]{\mkern3mu\overline{\mkern-3mu #1\mkern-3mu}\mkern3mu}
\usepackage{algorithm}
\usepackage{algcompatible}
\usepackage[noend]{algpseudocode}
\usepackage{cite}

\newcommand{\xx}{\mathbf{x}}
\newcommand{\uu}{\mathbf{u}}
\newcommand{\ww}{\mathbf{w}}
\newcommand{\bfeta}{\boldsymbol{\eta}}

\newcommand{\Phix}{\mathbf{\Phi}_x}
\newcommand{\Phiu}{\mathbf{\Phi}_u}
\newcommand{\KK}{\mathbf{K}}

\newcommand{\tildeww}{\widetilde{\mathbf{w}}}

\newcommand{\zero}{\mathbf{0}}
\newcommand{\diag}{\textrm{diag}}
\newcommand{\blkdiag}{\textrm{blkdiag}}

\newcommand{\cc}{\mathbf{c}}
\newcommand{\ddelta}{\boldsymbol{\delta}}
\newcommand{\hh}{\mathbf{h}}
\newcommand{\vv}{\mathbf{v}}

\newcommand{\ee}{\mathbf{x}_e}
\newcommand{\ue}{\mathbf{u}_e}

\usepackage{xcolor}

\allowdisplaybreaks

\newtheorem{theorem}{Theorem}

\newtheorem{remark}{Remark}
    
\newtheorem{corollary}{Corollary}
\newtheorem{proposition}{Proposition}    
\title{{\LARGE \bf {Safety Filter Design for Neural Network Systems via \\ Convex Optimization} }}

\author{Shaoru Chen$^\dagger$, Kong Yao Chee$^\dagger$, Nikolai Matni, M. Ani Hsieh, George J. Pappas
\thanks{Shaoru Chen is with Microsoft Research, 300 Lafayette Street, New York, NY, 10012, USA, {\tt shaoruchen@microsoft.com}. Kong Yao Chee, Nikolai Matni, M. Ani Hsieh, George J. Pappas are with the Department of Electrical and Systems Engineering, University of Pennsylvania, Philadelphia, PA, 19104, USA. {\tt \{ckongyao, nmatni, m.hsieh, pappasg\}@seas.upenn.edu.}  
\newline
$^\dagger$ The first two authors contributed equally and are co-first authors. 
}
}
 
\date{}

\begin{document}
\pagestyle{plain}
%\pagenumbering{arabic}
\maketitle

\begin{abstract}
With the increase in data availability, it has been widely demonstrated that neural networks (NN) can capture complex system dynamics precisely in a data-driven manner. However, the architectural complexity and nonlinearity of the NNs make it challenging to synthesize a provably safe controller. In this work, we propose a novel safety filter that relies on convex optimization to ensure safety for a NN system, subject to additive disturbances that are capable of capturing modeling errors. Our approach leverages tools from NN verification to over-approximate NN dynamics with a set of linear bounds, followed by an application of robust linear MPC to search for controllers that can guarantee robust constraint satisfaction. We demonstrate the efficacy of the proposed framework numerically on a nonlinear pendulum system. 
\end{abstract}

% introduction
\section{Introduction}
\label{sec:introduction}
With the rapid development in machine learning infrastructure, neural networks (NN) have been ubiquitously applied in the modeling of complex dynamical systems~\cite{ogunmolu2016nonlinear, chee2022knode}. Through a data collection and training procedure \cite{paszke2017automatic}, NNs can capture accurate representations of the system dynamics even in challenging scenarios where high-speed aerodynamic effects~\cite{bansal2016learning, bauersfeld2021neurobem, o2022neural} or contact-rich environments~\cite{yang2020data, williams2017information} are present. Moreover, NNs can be easily updated online as more data is collected, making them suitable for online tasks or modeling changing environments. For example, NN dynamical systems are widely used in model-based reinforcement learning~\cite{nagabandi2018neural} and learning-based adaptive control~\cite{o2022neural}.

However, applying NN dynamics brings about significant challenges in providing safety guarantees for the controlled system. Benefiting from the expressivity of NNs, we are meanwhile faced with the high nonlinearity and large scale of NNs since they are often overparameterized. The runtime assurance (RTA)~\cite{hobbs2023runtime} mechanism provides a practical and effective solution to guarantee the safety of complex dynamical systems by designing a safety filter that primarily focuses on enforcing safety constraints. Given a primary controller that aims to optimize performance, the safety filter monitors and modifies the output of the primary controller online to guarantee that only safe control inputs are applied. The safety filter allows the design of the safety and performance-based controllers to be decoupled and has found wide applications in safe learning-based control~\cite{emam2022safe, tearle2021predictive, alshiekh2018safe}.

In this work, we focus on the design of a predictive safety filter (PSF)~\cite{wabersich2021predictive} for uncertain NN dynamics. The PSF essentially follows a model predictive control (MPC) formulation with the nonlinear dynamics and constraints encoded in an optimization problem. Different from MPC, the PSF is less complex to solve since it does not consider any performance objectives~\cite{wabersich2021predictive}. Compared with the alternative safety filter construction schemes through control barrier functions (CBF)~\cite{ames2016control, ma2022learning} or Hamilton-Jacobi (HJ) reachability analysis~\cite{akametalu2014reachability, fisac2018general}, the PSF enjoys flexibility in handling dynamically changing NN models or model uncertainty bounds when updated online. We refer the interested readers to~\cite[Section 1 and 2]{wabersich2021predictive} for a detailed discussion of the PSF, CBF, and HJ reachability-based safety filters. 

\begin{figure}
    \centering
    \includegraphics[width = 0.9\columnwidth, trim = 0.0cm 0.8cm 0cm -0.8cm]{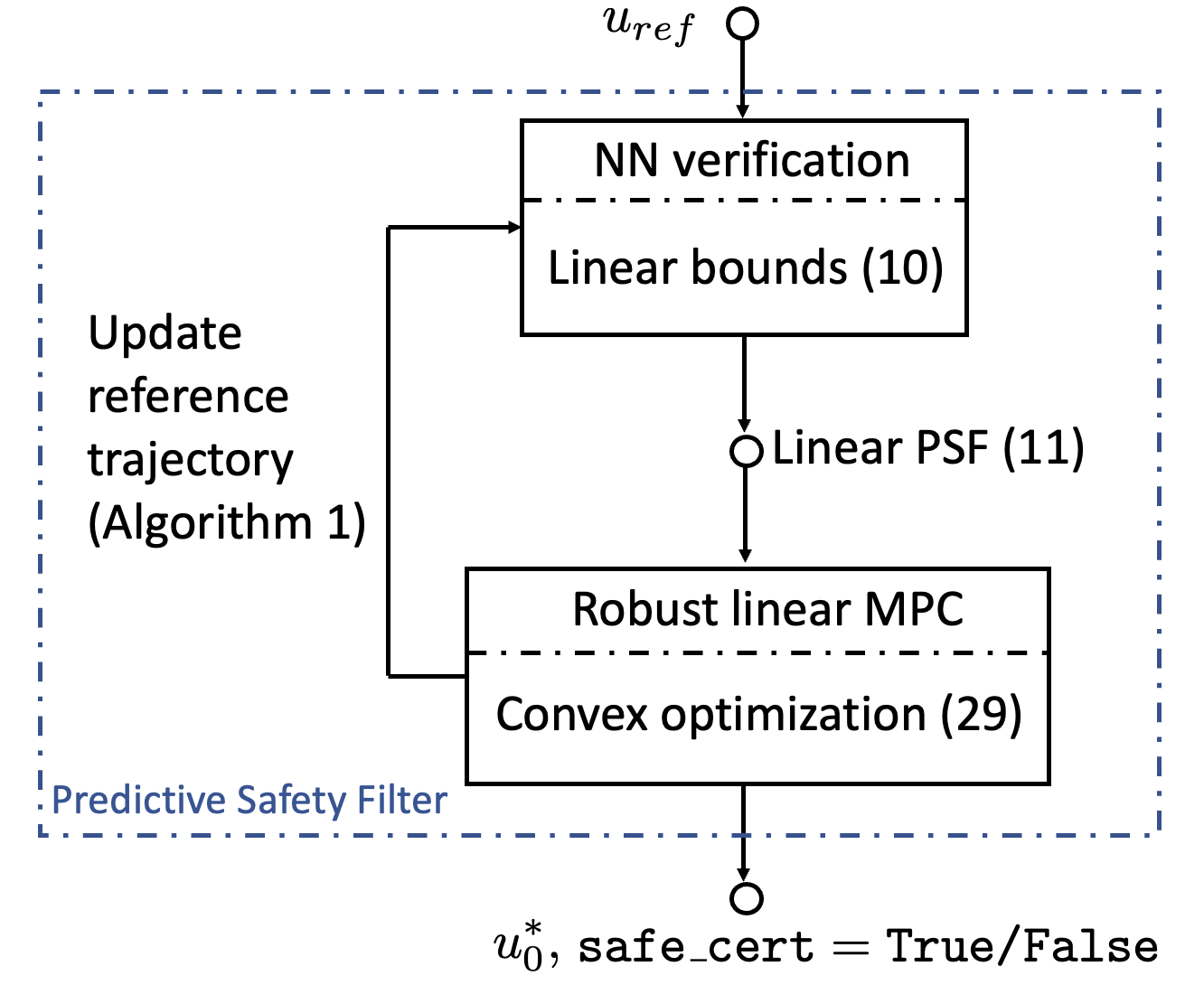}
    \caption{Pipeline of the proposed predictive safety filter for uncertain NN systems. }
    \label{fig:workflow}
\end{figure}

\textbf{Contributions}: In this work, we consider uncertain NN dynamics subject to bounded additive disturbances, where the disturbances can encapsulate the errors between the learned NN dynamics and the true system. Despite being highly expressive, the considered uncertain NN dynamics requires solving a robust optimization problem involving NN dynamical constraints online in the PSF. To resolve this computational challenge, we propose to apply NN verification tools~\cite{xu2020automatic} to abstract the NN dynamics locally as a linear uncertain system, thereby reducing the original PSF problem into one that is amenable to robust linear MPC and convex optimization. In particular, we adapt the SLS (System Level
Synthesis) MPC method~\cite{chen2022robust} to solve the resulting robust MPC
problem. A schematic of our pipeline is shown in Fig.~\ref{fig:workflow}. Soft constraints are used in robust linear MPC where slack variables denoting constraint violations are penalized. By applying a hierarchy of conservative function and model uncertainty approximations, we transform the original optimization problem into a convex one. A safety certificate for the uncertain NN dynamics over a finite horizon can then be provided when all slack variables are zero. Our contributions are summarized below. 
\begin{enumerate}
    \item Drawing tools from NN verification and robust linear MPC, we propose a novel predictive safety filter for uncertain NN dynamics through convex optimization. Importantly, the complexity of the convex optimization problem is independent of the NN size (i.e., width and depth of the NN). 
    \item Our PSF provides a safety certificate for the uncertain NN dynamics over a finite horizon when a certain numerical criterion is met by the convex optimization solutions. 
    % \item We demonstrate the effectiveness and scalability of the proposed PSF over a wide range of nonlinear systems. 
\end{enumerate}

\subsection{Related works}
% \textbf{Safety filter}

% \textbf{MPC of NN dynamics}

% \textbf{NN verification and robust MPC}

The problem of ensuring the safety of learning-based systems has received significant interest with a plethora of methods described in \cite{brunke2022safe}. Directly related to our work is the PSF developed in \cite{wabersich2021predictive}, which monitors and modifies a given control input by solving a predictive control problem online to guarantee the safety of the system. This formulation has been extended to the SLS setting~\cite{leeman2023predictive}, applied to racing cars~\cite{tearle2021predictive} and a soft-constrained variant is proposed in~\cite{wabersich2022predictive} to handle unexpected disturbances to the states. The PSF that we propose differs from those in the existing work in the following ways. First, we exploit the structure of the neural networks to extract linear bounds on the NN outputs using NN verification tools~\cite{zhang2018efficient,xu2020automatic}, simplifying the PSF formulation for NN dynamics. Second, our proposed pipeline circumvents the need to solve a robust non-convex optimization problem, even with the consideration of additive disturbances within the uncertain dynamics, as typical for nonlinear variants of SLS~\cite{leeman2023RobustSynthesis}. Unlike existing work in predictive control of NN dynamics~\cite{saviolo2022physics, spielberg2021neural,chee2022knode}, our work considers robust control of uncertain NN dynamics with a focus on obtaining formal safety guarantees. 

\textit{Notation}: $[N] $ denotes the set $\{0, 1, \dots, N\}$. $\mathcal{B}_p(z, r)$ denotes the $\ell_p$ ball centered at $z$ with radius $r$. We use $x_{0:T}$ to denote the sequence $\{x_0,\dotsc,x_T\}$. $\mathbf{0}$ denotes a vector of all zeros, and its dimension can be inferred from context. For a vector $d \in \mathbb{R}^n$, $S = \diag(d) \in \mathbb{R}^{n \times n}$ denotes a diagonal matrix with $d$ being the diagonal vector. For a sequence of matrices $S_1, \cdots, S_N$, $S = \text{blkdiag}(S_1, \cdots, S_N)$ denotes a block diagonal matrix whose diagonal blocks are $S_1, \cdots, S_N$ arranged in the order. We represent a linear, causal operator $\mathbf{R}$ defined over a horizon $T$ by the block-lower-triangular matrix
\begin{equation} 
	\mathbf{R} = \begin{bmatrix}
		R^{0,0} & \ & \ & \ \\
		R^{1,1} & R^{1,0} & \ & \ \\
		\vdots & \ddots & \ddots & \ \\
		R^{T,T} & \cdots &R^{T,1} & R^{T,0}
	\end{bmatrix}
\end{equation} 
where $R^{i,j} \in \mathbb{R}^{p \times q}$ is a matrix of compatible dimension. The set of such matrices is denoted by $\mathcal{L}_{TV}^{T, p \times q}$ and the superscript $T$ or $p \times q$ will be dropped when it is clear from the context. 

% problem formulation 
\section{Problem Formulation}
\label{sec:problem_formulation}
Consider the following discrete-time nonlinear system,
\begin{equation}\label{eq:dynamics}
	x(k+1) = Ax(k) + Bu(k) + f\left(x(k), u(k)\right) + w(k),
\end{equation}
where the vectors $x(k) \in \mathbb{R}^{n_x}$ and $u(k) \in \mathbb{R}^{n_u}$ are the state and control input. The vector $w(k)\in \mathbb{R}^{n_x}$ denotes the additive disturbances that can account for unknown effects from the environment or unmodeled dynamics. We assume that $w(k)$ is norm-bounded, $\textit{i.e.}$, $w(k) \in \mathcal{W} := \{ w \mid \lVert w \rVert_\infty \leq \sigma_w \}$. The system consists of a set of linear dynamics characterized by the matrices $A$ and $B$ and a set of nonlinear dynamics, $f(x(k), u(k))$. Specifically, these nonlinear dynamics are given by a NN $f:\mathbb{R}^{n_x + n_u} \mapsto \mathbb{R}^{n_x}$ with an arbitrary architecture.  While we discuss our proposed approach taking into account a general formulation of dynamics in \eqref{eq:dynamics}, the approach allows the matrices $A$ and $B$ to be zero and has the flexibility to account for time-varying dynamics $(A(k), B(k))$. The system \eqref{eq:dynamics} is required to satisfy the following constraints, 
\begin{equation} \label{eq:constraints}
	x(k) \in \mathcal{X}, \quad u(k) \in \mathcal{U},
\end{equation}
where $\mathcal{X} \subset \mathbb{R}^{n_x}$ and $\mathcal{U} \subset \mathbb{R}^{n_u}$ are polytopes. The state $x(k)$ and input $u(k)$ are considered \emph{safe} if they satisfy constraints~\eqref{eq:constraints}. 

% To achieve constraint satisfaction at all times $k \geq 0$, one approach is to consider a robust optimization problem, which involves solving a non-convex optimization problem that accounts for the dynamics in system \eqref{eq:dynamics}, and in particular, for all $w(k) \in \mathcal{W}$. However, this approach is computationally expensive and intractable in general. 

% In the case where there is an existing nominal control policy $\pi(x)$ for the system, an alternative and more practical approach is to design a \emph{predictive safety filter}. This filter modifies the control inputs computed from the nominal policy such that the system can be rendered safe.

% Due to the nonlinearities and architectural complexity of $f(x, u)$, guaranteeing constraint satisfaction of system~\eqref{eq:dynamics} for all time $k \geq 0$ is intractable in general. In this work, we approach this problem from a practical perspective. Instead of assuming the availability of a robust forward invariant set for system~\eqref{eq:dynamics} which is hard to find itself, we aim to develop a numerically scalable and efficient framework to control the system such that 
% %
% \begin{itemize}
%     \item the safety of the system is promoted, and
%     \item we can provide formal robust constraint satisfaction guarantees for the trajectory of system~\eqref{eq:dynamics} over a finite horizon whenever possible. 
% \end{itemize}
% %
% This is achieved by designing a predictive safety filter introduced below. 

\subsection{Predictive safety filter} 
We assume a primary controller $\pi(x)$ which aims to complete a task or achieve high performance is given for system~\eqref{eq:dynamics}. Following the runtime assurance scheme, the primary controller $\pi(x)$ is not guaranteed to be safe since its design may be decoupled from the safety requirements. To ensure constraint satisfaction of the closed-loop system, we aim to design a predictive safety filter that monitors and modifies the control input $\pi(x(k))$ given by the primary controller in a minimally invasive manner online. This is achieved by solving the following robust optimization problem at each time step $k$, 
\begin{equation} \label{eq:safety_filter}
\begin{aligned}
    \underset{u_{0:T-1}}{\textrm{minimize}} & \quad \lVert u_0 - \pi(x(k)) \rVert_2^2 \\
    \textrm{subject to} & \quad 	x_{t+1} = Ax_t + Bu_t + f(x_t, u_t) + w_t, \\
    & \quad x_t \in \mathcal{X}, u_t \in \mathcal{U}, \forall w_t \in \mathcal{W},\quad t \in [T], \\
    &\quad x_0 = x(k),
\end{aligned}
\end{equation}
where the vectors $x_t,\, u_t$ denote the predicted states and inputs over the horizon $T$, with $x(k)$ representing the current state of the system. We denote Problem~\eqref{eq:safety_filter} as the PSF problem and the sequence $u_{0:T-1}^*$ as the optimal solution of Problem~\eqref{eq:safety_filter}. When applied, the control inputs $u_{0:T-1}^*$ can guarantee the safety of the system for the next $T$ steps. In practice, the PSF problem~\eqref{eq:safety_filter} is solved recursively at each time step $k$, and the first optimal control input $u_0^*$ is applied to the system, analogous to an MPC scheme. 

\subsection{Challenges with the predictive safety filter}
While the solution to Problem~\eqref{eq:safety_filter} is able to provide safety guarantees, solving Problem~\eqref{eq:safety_filter} is a challenging task. Some potential issues include
\begin{enumerate}[(i)]
    \item it is well known in robust MPC that searching over open-loop control sequences $u_{0:T-1}$ can be overly conservative \cite{mayne2014model},
    \item the presence of the NN dynamics $f(x, u)$ makes solving the PSF computationally challenging,
    \item the safety certificate of the solution $u_{0:T-1}^*$ is not available until convergence is reached, and
    % comment: add a reference to robust positive invariant set later.
    \item without the availability of a robust forward invariant set, attempting to solve the PSF may result in infeasibility.
\end{enumerate}

To handle all the aforementioned issues, we combine NN verification tools \cite{xu2020automatic} and robust MPC \cite{chen2022robust}. Our solution consists of two steps. First, we generate local linear bounds for the NN dynamics $f(x, u)$ using tools from NN verification. Next, we apply robust linear MPC to synthesize a state feedback control policy that guarantees robust constraint satisfaction for the system. This combined procedure provides a powerful simplification of the PSF problem and resolves issues (i) to (iii). To address the issue (iv), we introduce soft constraints in our formulation. This provides formal safety guarantees for the system when the slack variables are zero. We describe these two steps in Sections~\ref{sec:nnv_bounds} and~\ref{sec:robust_mpc}, respectively. Section~\ref{sec:simulation} demonstrates our method numerically, and Section~\ref{sec:conclusion} concludes the paper. 

% Note that in the work, we focus on evaluating the practical performance of the proposed PSF rather than providing rigorous safety guarantees \emph{at all times}. The latter requires the availability of a locally safe controller and a forward invariant set, which are challenging to find even for static NN dynamics. We leave the synthesis of the forward invariant terminal set for NN dynamics and its integration into the PSF for future research.  

\begin{remark}
{To ensure the safety constraints are satisfied \emph{at all times} or guarantee recursive feasibility of the PSF~\eqref{eq:safety_filter}, a local forward invariant set for the nonlinear system~\eqref{eq:dynamics} is required, which is generally challenging to find. In this work, we do not assume the availability of such a forward invariant set and use slack variables as numerical certificates of safety. We leave the synthesis of the forward invariant terminal set for NN dynamics and its integration into the PSF as part of our future work.  }
\end{remark}

% On the other hand, while NN verification can certify the safety (in terms of constraint satisfcation)  of the closed-loop system under the nominal policy $\pi(x)$, it does not provide a correction of the control input $\pi(x)$ in case of verification failure. In this work, we propose an efficient, convex optimization based approach to design a safety filter in a conservative manner. This allows us to obtain a safe control input and a numerical safety ceritifcate more efficiently than solving Problem~\eqref{eq:robust_optimization}. 

%Synthesizing a controller for system~\eqref{eq:dynamics} such that constraints~\eqref{eq:constraints} are satisfied is challenging. Data-driven methods such as~\cite{} have been proposed to synthesize a controller $u_t = \pi(x_t)$, but formal guarantees on $\pi(\cdot)$ are unavailable. 

% neural network verification 
\section{Neural network verification bounds}
\label{sec:nnv_bounds}
In this section, we demonstrate how tools from NN verification can be utilized to over-approximate the NN dynamics with a linear time-varying (LTV) representation. This enables us to conservatively transform the PSF problem into a robust convex optimization problem, which is simpler to solve. 

% \subsection{Linear bounds for NNs}
Given a bounded input set, the linear relaxation-based perturbation analysis (LiRPA)~\cite{xu2020automatic} is an efficient method to synthesize linear lower and upper bounds for the outputs of a NN with a general architecture. The bounds computed from this method are described in the following theorem. 
\begin{theorem} (rephrasing \cite[Theorem 3.2]{zhang2018efficient}) \label{thm:linear_bounds}
Given a NN $f(z): \mathbb{R}^{n_0} \mapsto \mathbb{R}^{n_L}$, there exist two explicit linear functions $f_U:  \mathbb{R}^{n_0} \mapsto \mathbb{R}^{n_L}$ and $f_L: \mathbb{R}^{n_0} \mapsto \mathbb{R}^{n_L}$ such that for all $z \in \mathcal{B}_p(z_0, r)$, we have 
\begin{equation} \label{eq:crown_bounds}
    f_L(z)= A_L z + b_L \leq f(z) \leq A_U z + b_U = f_U(z),
\end{equation}
where the inequalities are applied component-wise.
\end{theorem}

The parameters $A_L, A_U, b_L, b_U$ are derived from the weights, biases and activation functions of the NN. In this paper, we choose $p := \infty$ such that $\mathcal{B}_\infty(z, r)$ is polyhedral. The bounds~\eqref{eq:crown_bounds} are computed using closed-form updates with a computational complexity polynomial in the number of neurons~\cite{zhang2018efficient}. This allows the method to scale well to deep networks. However, if the NN is deep or if the input domain $\mathcal{B}_\infty(z, r)$ is large, the computed bounds tend to be loose. Motivated by this observation, we propose to extract a set of local linear bounds along a \emph{reference trajectory}. %obtained using the nominal control policy $\pi(x)$. %This allows us to search for robust controllers locally, while maintaining constraint satisfaction. 
Specifically, at every time step $k$, we construct a reference trajectory given by the sequences of reference states $\mathbf{\hat{x}} := \hat{x}_{0:T}$ and control inputs $\mathbf{\hat{u}} := \hat{u}_{0:T-1}$ where
\begin{equation} \label{eq:ref_traj}
	\begin{aligned}
	& \hat{x}_{t+1}  = A\hat{x}_t + B \hat{u}_t + f(\hat{x}_t, \hat{u}_t),\quad t \in [T-1], \\
	& \hat{x}_0= x(k).
	\end{aligned}
\end{equation}
The reference control inputs $\mathbf{\hat{u}}$ can be obtained, e.g., by rolling out the nominal NN dynamics following the primary policy $\pi(\cdot)$. By denoting ${z}_t := [{x}_t^\top \ {u}_t^\top]^\top$, $\hat{z}_t := [\hat{x}_t^\top \ \hat{u}_t^\top]^\top$ and $r_t$ to be the radius of the $\ell_{\infty}$ ball around $\hat{z}_t$, we apply Theorem~\ref{thm:linear_bounds} to get the following bounds for the NN dynamics along the reference trajectory,
\begin{equation} \label{eq:linear_bounds_direct}
\begin{bmatrix} \ubar{G}_t^x & \ubar{G}_t^u \end{bmatrix} z_t + \underline{g}_t  \leq f(x_t, u_t) \leq \begin{bmatrix}\obar{G}_t^x & \obar{G}_t^u  \end{bmatrix} z_t + \overline{g}_t,
\end{equation}
for all $(x_t, u_t) \in \mathcal{B}_\infty(\hat{z}_t, r_t)$. In other words, the NN dynamics $f(x_t, u_t)$ is over-approximated with a set of linear lower and upper bounds. The ball $\mathcal{B}_\infty(\hat{z}_t, r_t)$ is referred to as the \emph{trust region} in which the bounds \eqref{eq:linear_bounds_direct} are valid. %This characterization enables the application of robust linear MPC to search for a robustly feasible control sequence $\tilde{u}_{0:T-1}^*$ within the vicinity of the nominal trajectory $(\mathbf{\hat{x}}, \mathbf{\hat{u}})$. 
%We observe that there is some flexibility in the selection of the nominal control inputs $\mathbf{\hat{u}}$. For instance, they can be generated by applying the nominal control policy recursively as $\hat{u}_t = \pi(\hat{x}_t)$ or can be obtained from an external control policy. With this flexibility, we treat the nominal trajectory $(\mathbf{\hat{x}}, \mathbf{\hat{u}})$ as hyperparameters and provide a method on how it can be chosen and updated in Section~\ref{sec:trust_region}.

To reduce conservatism in the formulation of the filter within the robust MPC framework, we integrate the bounds into the linear dynamics of the system. 
%Although we can bound the nonlinear NN dynamics by linear bounds in~\eqref{eq:linear_bounds_direct}, it is overly conservative to treat $f(x_t, u_t)$ as uncertainty. 
Specifically, using the bounds in \eqref{eq:linear_bounds_direct}, we define
\begin{equation} \label{eq:uncertain_dyn}
f_d(x_t, u_t) := f(x_t, u_t) - \left(\tilde{A}_t x_t + \tilde{B}_t u_t + \tilde{c}_t\right),
\end{equation}
where 
\begin{equation} \label{eq:mean_bounds}
\tilde{A}_t := \frac{\ubar{G}_t^x  + \obar{G}_t^x }{2}, \; \tilde{B}_t := \frac{\ubar{G}_t^u  + \obar{G}_t^u }{2}, \; \tilde{c}_t := \frac{\underline{g}_t + \overline{g}_t}{2}
\end{equation}
denote the means of the linear bounds. 
% These uncertain dynamics enjoy the following bounds.
\begin{corollary} \label{cor:extraction}
Given the bounds in \eqref{eq:linear_bounds_direct}, for every $z_t:=[x_t^{\top}\,u_t^{\top}]^{\top}\in \mathcal{B}_\infty(\hat{z}_t, r_t)$, the dynamics $f_d(x_t, u_t)$ in \eqref{eq:uncertain_dyn} have the following bounds,
\begin{equation} \label{eq:linear_bounds_shift}
\begin{bmatrix} \ubar{D}_t^x & \ubar{D}_t^u \end{bmatrix} z_t + \underline{d}_t \leq f_d(x_t, u_t) \leq \begin{bmatrix} \obar{D}_t^x & \obar{D}_t^u \end{bmatrix} z_t +\overline{d}_t,
\end{equation}
where 
\begin{equation*}
\small \begin{split}
\ubar{D}_t^x = \frac{\ubar{G}_t^x  - \obar{G}_t^x }{2} &= -\obar{D}_t^x,\;\;
\ubar{D}_t^u = \frac{\ubar{G}_t^u  - \obar{G}_t^u }{2} = -\obar{D}_t^u,\\
&\underline{d}_t = \frac{\underline{g}_t - \overline{g}_t}{2} = -\overline{d}_t.
\end{split}
\end{equation*}
\end{corollary}

It is important to note that although the NN dynamics $f(x_t, u_t)$ can have large values within the trust region $ \mathcal{B}_\infty(\hat{z}_t, r_t)$, the dynamics $f_d(x_t, u_t)$ tends to be small in magnitude and can be treated as disturbances to the system. With the extraction of $f_d(x_t, u_t)$, we obtain an LTV reformulation of the PSF problem, referred to as the \emph{linear PSF} problem,
\begin{equation} \label{eq:robust_filter}
\begin{aligned}
	\underset{u_{0:T-1}}{\textrm{minimize}} & \quad \lVert u_0 - \pi(x(k)) \rVert_2^2 \\
	\textrm{subject to} & \quad x_{t+1} = A_t x_t + B_t u_t + c_t+  \Delta_t(x_t, u_t) + w_t, \\
  & \quad (x_t, u_t) \in  \mathcal{B}_\infty \left(\hat{z}_t, r_t\right),\quad t \in [T-1],\\
	& \quad x_t \in \mathcal{X},\, u_t \in \mathcal{U},\quad t \in [T], \\
 & \quad \forall \Delta_t(\cdot) \in \mathcal{P}_t,\, w_t \in \mathcal{W},\quad t \in [T], \\
	& \quad x_0 = x(k),
\end{aligned}
\end{equation}
% Comment: we need to specify p=inf somewhere. 
where the means in \eqref{eq:mean_bounds} are merged into the linear dynamics of the system with the definition of the following time-varying system parameters~\footnote{When a time-varying dynamics $(A(k), B(k))$ is considered in~\eqref{eq:dynamics}, we can replace $(A, B)$ by their time-varying counterparts in the definitions of $(A_t, B_t)$. }, 
\begin{equation*}
A_t := A + \tilde{A}_t,\, B_t := B + \tilde{B}_t,\,c_t := \tilde{c}_t.
\end{equation*}
The uncertainty set $\mathcal{P}_t$ is given as
\begin{equation} \label{eq:uncertainty_set}
\mathcal{P}_t := \Bigg \{\hspace{-0.1cm} \Delta_t(x_t, u_t) \Bigg|\hspace{-0.15cm} \begin{array}{l}
\Delta_t(x_t, u_t) \geq \begin{bmatrix} \ubar{D}_t^x & \ubar{D}_t^u \end{bmatrix} z_t + \underline{d}_t   \\
\Delta_t(x_t, u_t) \leq \begin{bmatrix} \obar{D}_t^x & \obar{D}_t^u \end{bmatrix} z_t +\overline{d}_t
\end{array}\hspace{-0.2cm}\Bigg\},
\end{equation}
using the bounds obtained from Corollary \ref{cor:extraction}. It immediately follows that $f_d(x_t,u_t)$ corresponds to a realization of $\Delta_t(x_t, u_t)$.

A few remarks about the linear PSF problem are in order. First, the solution of the linear PSF problem is dependent on the centers $\mathbf{\hat{z}} = \hat{z}_{0:T}$ and radius $r_t$ of the trust regions. The centers play an important role when the reference trajectory lies near or beyond the boundaries of the constraint set. In this case, the reference trajectory should be shifted towards the constraint set and this is done by adjusting the centers of these trust regions. Next, based on how the linear bounds in~\eqref{eq:uncertainty_set} are computed, there is a trade-off in the size of the radius $r_t$. A small radius ensures that the computed bounds are accurate, but it limits the range in which the centers $\mathbf{\hat{z}}$ can be updated at each iteration. On the other hand, a large radius provides more flexibility in updating the reference trajectory, but the bounds can be overly conservative. Lastly, any feasible solution $u^*_{0:T-1}$ to the linear PSF problem~\eqref{eq:robust_filter} is also feasible for the PSF problem~\eqref{eq:safety_filter}.
 
With these considerations, we devise a method such that the trust regions $\mathcal{B}_\infty(\hat{z}_t,\, r_t)$ can be updated online. Details of this update are given in Section \ref{sec:trust_region_update}. In Section~\ref{sec:sls_mpc} to~\ref{sec:convex_PSF}, we first describe how to solve the linear PSF problem through robust linear MPC.

% robust MPC
\section{Robust Linear MPC}
\label{sec:robust_mpc}
Compared with the PSF problem~\eqref{eq:safety_filter}, the linear PSF problem~\eqref{eq:robust_filter} only involves uncertain linear dynamics. However, solving it can still be a challenge and a conservative approach may not succeed. Since optimizing over open-loop control sequences is conservative in robust MPC, we consider optimizing over state-feedback controllers $u_t = \mu_t(x_{0:t})$ instead. To achieve this, we apply an extension of the SLS MPC algorithm in~\cite{chen2022robust} which is shown to enjoy outstanding tightness among the existing robust linear MPC methods.

\subsection{Overview of SLS MPC}
\label{sec:sls_mpc}
In SLS MPC, we consider the following uncertain linear time-varying system,
\vspace{-0.1cm}
\begin{equation}
\vspace{-0.1cm} \label{eq:sls_dynamics}
    x_{t+1} = A_t x_t + B_t u_t + \xi_t 
\end{equation}
where the time-varying matrices that represent the nominal dynamics $(A_t, B_t)$ are known and $\xi_t$ denotes the lumped uncertainty, which will be used to capture the effects of uncertainty in the dynamics. 

Consider the dynamics~\eqref{eq:sls_dynamics} over a horizon $T$. We define the following variables, which are concatenations of the variables in~\eqref{eq:sls_dynamics} over the horizon $T$,
\begin{equation} \label{eq:signal_def}
\begin{aligned}
&\xx := [x_0^\top \ \cdots \ x_T^\top]^\top, \quad\quad \ \uu := [u_0^\top \ \cdots \ u_T^\top]^\top, \\ 
&\boldsymbol{\xi} := [x_0^\top \ \xi_0^\top \ \cdots \ \xi_{T-1}^\top]^\top, 
\end{aligned}
\end{equation}
and these concatenated system matrices,
\vspace{-0.1cm}
\begin{equation*} 
\vspace{-0.1cm}
    \mathbf{{A}} := \blkdiag\left({A}_0, \cdots, {A}_{T}\right), \, \mathbf{{B}} := \blkdiag\left({B}_0, \cdots, {B}_T\right).
\end{equation*}
% comment (KY): If x_0 is in \xi, wouldn't (18) be inconsistent?
%Note that $\xx, \uu, \boldsymbol{\xi}$ can be interpreted as finite-horizon signals, and the initial state $x_0$ is included as the first component of $\boldsymbol{\xi}$.
We define $Z$ as the block-downshift operator with the first block sub-diagonal filled with identity matrices and zeros everywhere else. The dynamics~\eqref{eq:sls_dynamics} over the horizon $T$ can then be compactly written as,
\vspace{-0.1cm}
\begin{equation}
\vspace{-0.1cm}
\label{eq:open_sls_dynamics}
    \xx = Z \mathbf{A} \xx + Z \mathbf{B} \uu + \boldsymbol{\xi},
\end{equation}
Next, we consider a LTV state feedback controller $u_t = \sum_{i=0}^t K^{t, t-i} x_i$, compactly given as $\uu = \KK \xx$, $\KK \in \mathcal{L}_{TV}^{T, n_u \times n_x}$. 
Plugging $\uu$ into~\eqref{eq:open_sls_dynamics} gives the following \emph{system responses} $\{ \Phix, \Phiu \}$, mapping $\boldsymbol{\xi}$ to the closed-loop states and inputs $(\xx,\uu)$,
\vspace{-0.1cm}
\begin{equation} \label{eq:closed_sls_dynamics}
\begin{bmatrix} \Phix \\ \Phiu \end{bmatrix} := \begin{bmatrix}
(I - Z (\mathbf{A}+ \mathbf{B} \KK))^{-1} \\
\KK (I - Z (\mathbf{A} + \mathbf{B} \KK))^{-1}
\end{bmatrix}.
\end{equation}
%Due to the block-downshift operator $Z$, the matrix inversion in~\eqref{eq:closed_sls_dynamics} is well-defined and $\Phix, \Phiu \in \mathcal{L}_{TV}^T$.

The following theorem establishes the connection between $\{\Phix, \Phiu\}$ and state feedback controllers. 
\begin{theorem}~\cite[Theorem 2.1]{anderson2019system}
\label{thm:SLS}
Over the horizon $t \in [T]$, for the dynamics~\eqref{eq:sls_dynamics} with the LTV state feedback controller $\uu = \KK \xx,\,\KK \in \mathcal{L}_{TV}^{T, n_u \times n_x}$, we have:
\begin{enumerate}
    \item The affine subspace defined by 
    \begin{equation} \label{eq:affine_constr}
    \begin{bmatrix} I - Z \mathbf{A} & -Z \mathbf{B} \end{bmatrix} 
    \begin{bmatrix} \Phix \\ \Phiu \end{bmatrix} = I, \ \Phix, \Phiu \in \mathcal{L}_{TV}^T
    \end{equation}
    parameterizes all possible system responses~\eqref{eq:closed_sls_dynamics}.
    \item For any $\lbrace \Phix, \Phiu \rbrace \in \mathcal{L}_{TV}^T$ satisfying~\eqref{eq:affine_constr}, the controller gain $\KK = \Phiu \Phix^{-1} \in \mathcal{L}_{TV}^T$ achieves the desired responses~\eqref{eq:closed_sls_dynamics}. 
\end{enumerate}
\end{theorem}
The system responses explicitly characterize the effects of the lumped uncertainty $\boldsymbol{\xi}$ on $(\xx, \uu)$. In a previous work\cite{chen2022robust}, a system subjected to polytopic uncertainties and additive disturbances is considered, \textit{i.e.}, $\xi_t := \Delta_A x_t + \Delta_B u_t + w_t$ where the parameters $(\Delta_A, \Delta_B)$ belong to a polytopic set. Interested readers are referred to~\cite{chen2022robust} for more details on SLS MPC. 

Despite its outstanding performance in conservatism reduction compared to existing robust MPC methods~\cite{chen2022robust}, applying SLS MPC directly to solve the linear PSF comes with these two challenges; (i)
% \begin{enumerate}
    the uncertainty set $\mathcal{P}_t$ is defined through affine inequalities and converting $\mathcal{P}_t$ into a vertex representation, which is amenable to existing robust MPC methods~\cite{langson2004robust, kohler2019linear, fleming2014robust, bujarbaruah2022robust}, requires $2^{n_x}$ vertices, 
    (ii) applying SLS MPC requires merging the constant $c_t$ into the lumped uncertainty $\xi_t$ which causes the bounds on $\boldsymbol{\xi}$ to be overly conservative. 
% \end{enumerate}
In the following subsections, we describe an extension to SLS MPC to address these two challenges. 

\subsection{Controller parameterization}
For the uncertain linear dynamics
\vspace{-0.1cm}
\begin{equation} \label{eq:uncertain_linear_dynamics}
\vspace{-0.1cm}
    x_{t+1} = A_t x_t + B_t u_t + c_t+  \Delta_t(x_t, u_t) + w_t
\end{equation}
stated in~\eqref{eq:robust_filter}, we define $\eta_t:=\Delta_t(x_t, u_t) + w_t$ as the lumped uncertainty. Instead of treating $\xi_t = c_t + \eta_t$ in~\eqref{eq:sls_dynamics}, we decompose~\eqref{eq:uncertain_linear_dynamics} into a set of nominal and error dynamics.

First, in addition to~\eqref{eq:signal_def}, we concatenate these variables over the horizon $T$,
\vspace{-0.05cm}
\begin{equation} \label{eq:def_eta_w}
\begin{aligned}
&\bfeta := [\zero^\top \ \eta_0^\top \ \cdots \ \eta_{T-1}^\top]^\top, \;  \ww := [\zero^\top \ w_0^\top \ \cdots \ w_{T-1}^\top]^\top, \\
& \cc := [\zero^\top c_0^\top \ \cdots \ c_{T-1}^\top]^\top, \quad \ddelta_{x_0} := [x_0^\top \ \zero^\top \ \cdots \ \zero^\top]^\top.
\end{aligned}
\end{equation}
We define the nominal and error states $\{\hh,\ee\}$ and control inputs $\{\vv,\ue\}$ as
\vspace{-0.1cm}
\begin{equation*}
\begin{split}
    \hh &:= [h_0^\top \ \cdots \ h_T^\top]^\top, \quad \vv := [v_0^\top \ \cdots \ v_T^\top]^\top,\\
    \ee &:= [x_{e,0}^\top \ \cdots \ x_{e,T}^\top]^\top = \xx - \hh, \\
    \ue &:= [u_{e,0}^\top \ \cdots \ u_{e,T}^\top]^\top = \uu - \vv,
\end{split}
\end{equation*}
with the nominal and error dynamics as
\vspace{-0.1cm}
\begin{subequations}
\begin{align}
    \hh &= Z (\mathbf{A} \hh + \mathbf{B} \vv) + \cc + \ddelta_{x_0}, \label{eq:nominal_dynamics}\\ 
     \ee &= Z (\mathbf{A} \ee + \mathbf{B} \ue) + \bfeta. \label{eq:error_dynamics}
\end{align}
\end{subequations}
It is important to note that \eqref{eq:error_dynamics} conforms with~\eqref{eq:sls_dynamics}. A LTV state feedback controller $\KK \in \mathcal{L}_{TV}^{T,n_u \times n_x}$ is then applied to control the error states. The overall controller for \eqref{eq:uncertain_linear_dynamics} is given by $\uu = \KK \ee + \vv = \KK (\xx - \hh) + \vv$.

\subsection{Lumped uncertainty over-approximation}
% For a set of control parameters $\{\KK, \hh, \vv\}$, we denote the reachable set of $\bfeta$ as 
% \begin{equation*}
% \begin{aligned}
% \mathcal{R}(\bfeta; \KK, \hh, \vv) := \{\bfeta \mid & \eta_t = \Delta_t(x_t, u_t) + w_t, \\ 
% & \Delta_t(\cdot) \in \mathcal{P}_t, t \in [T-1]\}.
% \end{aligned}
% \end{equation*}
For the lumped uncertainty $\bfeta$, its dependence on $\xx$, $\uu$ and $\ww$ complicates the design of the robust controller. As in SLS MPC, the approach is to over-approximate $\bfeta$ by an independent, filtered virtual disturbance signal $\mathbf{\Psi} \tildeww$, where
$
 \tildeww = [\zero^\top \ \tilde{w}_0^\top \ \cdots \ \tilde{w}_{T-1}^\top]^\top, \ \lVert \tilde{w}_t \rVert_\infty \leq 1.
$
The matrix $\mathbf{\Psi} \in \mathcal{L}_{TV}^{T, n_x \times n_x}$ is a filter operating on the finite-horizon virtual disturbance signal $\tildeww$, with its diagonal blocks of $\mathbf{\Psi}$ structured as 
$
% \begin{aligned}
\Psi^{0,0} = I, \Psi^{t,0} = \diag(\psi_{t-1})
$ with
$
\psi_{t-1}\in \mathbb{R}^{n_x}, \psi_{t-1} > 0, \, t = 1, \dots, T.
% \end{aligned}
$

We define $\widetilde{\mathcal{W}} = \left\{ \tildeww \mid \lVert \tilde{w}_t \rVert_\infty \leq 1, t \in [T-1] \right\}$ as the set of admissible virtual disturbances. Since $\tilde{w}_t$ are unit norm-bounded, we tune the filter $\mathbf{\Psi}$ to change the reachable set of $\mathbf{\Psi} \tildeww$, defined as 
$
    \mathcal{R}(\mathbf{\Psi} \tildeww) := \{  \boldsymbol{\zeta}  \mid \boldsymbol{\zeta} = \mathbf{\Psi} \tildeww, \tildeww \in \widetilde{\mathcal{W}} \},
$
$\boldsymbol{\zeta} := [\zero^\top \ \zeta_0^\top \ \cdots \ \zeta_{T-1}^\top]^\top$. 

Our goal is to find sufficient conditions on the control parameters $\{\KK, \hh, \vv\}$ and $\mathbf{\Psi}$ such that the reachable set of $\bfeta$, denoted by $\mathcal{R}\left(\bfeta;\, \KK, \hh, \vv\right):= \left\{\bfeta \mid \eta_t = \Delta_t(x_t, u_t) + w_t,
\Delta_t \in \mathcal{P}_t,\, t \in [T-1]\right\}$, is a subset of the reachable set of $\mathbf{\Psi} \tildeww$. 
%Before we derive the sufficient condition for~\eqref{eq:over_approximation} to hold, the parameterization of the $\mathbf{\Psi}$ needs to be clarified. In addition to 
% Essentially, the diagonal blocks $\Psi^{t,0}$ of $\mathbf{\Psi}$ are diagonal matrices, and the reachable set of $\Psi^{t,0} \tilde{w}_{t-1} = \diag(\psi_{t-1}) \tilde{w}_{t-1}$ is a hyperrectangle. 
The following proposition provides these sufficient conditions and the proof is given in Appendix~\ref{app:proof}. 
%The diagonal matrix parameterization of $\Psi^{t,0}$ allows formulating the sufficient condition as convex constraints. 

\begin{proposition} \label{prop:sufficient}
Let $e_i \in \mathbb{R}^{n_x}$ denote the $i$-th standard basis for $i = 1, \dots, n_x$ and $y_t = [h_t^\top \ v_t^\top]^\top$, $\Phi^{t,t-i} := \begin{bmatrix} \Phi_x^{t,t-i\top} & \Phi_u^{{t,t-i}\top}\end{bmatrix}^{\top}$ for $t \in [T-1]$. The following constraints:
\begin{equation} \label{eq:affine_scaled}
\begin{bmatrix} I - Z \mathbf{A} & -Z \mathbf{B} \end{bmatrix} 
\begin{bmatrix} \Phix \\ \Phiu \end{bmatrix} = \mathbf{\Psi}, \ \Phix, \Phiu \in \mathcal{L}_{TV}^T,
\end{equation}
and 
\begin{subequations} \label{eq:over_approx_constr}
\begin{align}
&\sigma_w + e_i^\top ([\obar{D}_{0}^x \; \obar{D}_{0}^u] y_0 + \overline{d}_{0}) \leq \psi_{0,i}, \label{eq:over_approx_0}\\
&\sigma_w - e_i^\top ([\ubar{D}_{0}^x \; \ubar{D}_{0}^u] y_0 + \underline{d}_{0}) \leq \psi_{0,i}, \label{eq:over_approx_1} \\
&\sigma_w + e_i^\top \Big([\obar{D}_t^x \; \obar{D}_t^u] y_t + \overline{d}_t \Big) + \nonumber \\
& \sum_{i=1}^t \left\lVert e_i^\top\Big([\obar{D}_t^x \; \obar{D}_t^u] \Phi^{t,t-i} -\Psi^{t+1,t+1-i}\Big) \right\rVert_1 \leq \psi_{t,i}, \label{eq:over_approx_2} \\
&\sigma_w - e_i^\top \Big([\ubar{D}_t^x \; \ubar{D}_t^u] y_t + \underline{d}_t \Big) + \nonumber\\
& \sum_{i=1}^t \left\lVert e_i^\top\Big([\ubar{D}_t^x \; \ubar{D}_t^u] \Phi^{t,t-i} -\Psi^{t+1,t+1-i}\Big) \right\rVert_1 \leq \psi_{t,i},  \label{eq:over_approx_3} \\
& i \in [n_x], \quad  t = 1, \cdots, T-1, \nonumber
\end{align}
\end{subequations}
% \begin{equation} \label{eq:over_approx_constr}
% \begin{aligned}
% &\sigma_w + e_i^\top ([\obar{D}_{0}^x \; \obar{D}_{0}^u] y_0 + \overline{d}_{0}) \leq \psi_{0,i},\\
% &\sigma_w - e_i^\top ([\ubar{D}_{0}^x \; \ubar{D}_{0}^u] y_0 + \underline{d}_{0}) \leq \psi_{0,i}, \\
% &\sigma_w + e_i^\top \Big([\obar{D}_t^x \; \obar{D}_t^u] y_t + \overline{d}_t \Big) +\\
% & \sum_{i=1}^t \left\lVert e_i^\top\Big([\obar{D}_t^x \; \obar{D}_t^u] \Phi^{t,t-i} -\Psi^{t+1,t+1-i}\Big) \right\rVert_1 \leq \psi_{t,i}, \\
% &\sigma_w - e_i^\top \Big([\ubar{D}_t^x \; \ubar{D}_t^u] y_t + \underline{d}_t \Big) +\\
% & \sum_{i=1}^t \left\lVert e_i^\top\Big([\ubar{D}_t^x \; \ubar{D}_t^u] \Phi^{t,t-i} -\Psi^{t+1,t+1-i}\Big) \right\rVert_1 \leq \psi_{t,i}, \\
% & i \in [n_x], \quad  t = 1, \cdots, T-1,
% \end{aligned}
% \end{equation}
guarantee that $\mathcal{R}(\bfeta; \KK, \hh, \vv) \subseteq \mathcal{R}(\mathbf{\Psi} \tildeww)$ holds.
\end{proposition}

% With these sufficient conditions, $\mathbf{\Psi} \tildeww$ over-approximates $\bfeta$ and we can solve the linear PSF \eqref{eq:robust_filter} using $\mathbf{\Psi} \tildeww$ to achieve robust constraint satisfaction guarantees. 

\subsection{Convex formulation of the linear PSF}
\label{sec:convex_PSF}
With the constraints~\eqref{eq:affine_scaled} and~\eqref{eq:over_approx_constr}, we have for any realization of $\bfeta$, there exists $\tildeww \in \widetilde{\mathcal{W}}$ such that $\bfeta = \mathbf{\Psi} \tildeww$. Therefore, we can represent $\bfeta$ as $\mathbf{\Psi}\tildeww$ and write the error dynamics~\eqref{eq:error_dynamics} as
\begin{equation} \label{eq:error_dynamics_tildeww}
    \ee = Z (\mathbf{A} \ee + \mathbf{B} \ue) + \mathbf{\Psi} \tildeww.
\end{equation}
By~\cite[Corollary 1]{chen2022robust}, we have constraint~\eqref{eq:affine_scaled} parameterizes all system responses $\ee = \Phix \tildeww, \ue = \Phiu \tildeww$ of system~\eqref{eq:error_dynamics_tildeww} under the controller $\ue= \KK \ee$. Then, the closed-loop states and control inputs of system~\eqref{eq:uncertain_linear_dynamics} are given by
% it suffices to consider the surrogate dynamics
% \begin{equation}\label{eq:surrogate}
%     \ee = Z (\mathbf{A} \ee + \mathbf{B} \ue) + \mathbf{\Psi} \tildeww
% \end{equation}
% in place of~\eqref{eq:error_dynamics} to guarantee robust constraint satisfaction of a controller $\{\KK, \hh, \vv \}$. By~\cite[Corollary 1]{chen2022robust}, the affine constraint~\eqref{eq:affine_scaled} parameterizes all system responses mapping $\tildeww$ to $(\mathbf{x}_e, \mathbf{u}_e)$, i.e., $\xx_e = \Phix \tildeww, \uu_e = \Phiu \tildeww$, for the surrogate system under the controller $\uu_e = \KK \xx_e$. 
\vspace{-0.1cm}
\begin{equation}\label{eq:closed_loop_states_inputs}
\vspace{-0.1cm}
\xx = \hh + \Phix \tildeww, \quad \uu = \vv + \Phiu \tildeww.
\end{equation}
To guarantee robust constraint satisfaction of the controller $\uu = \KK(\xx - \hh) + \vv$, we tighten the constraints in the linear PSF \eqref{eq:robust_filter}. Consider the state constraint $x_t \in \mathcal{X}$ as an example. The constraints are represented by a polyhedral set, $\mathcal{X} = \lbrace x \in \mathbb{R}^{n_x} \mid F_x x \leq b_x \rbrace$, where $F_x \in \mathbb{R}^{n_\mathcal{X} \times n_x}$, $b_x := [b_1^x, \dotsc,b_{n_{\mathcal{X}}}^x]^{\top} \in \mathbb{R}^{n_\mathcal{X}}$, and $\{a_j^{x \top}\}_{j=1}^{n_{\mathcal{X}}}$ denote the rows of $F_x$. This implies that $(a_j^x,b_j^x)$ denotes the $j$-th set of linear constraint parameters in $\mathcal{X}$. From~\eqref{eq:closed_loop_states_inputs}, we have $x_t = h_t + \sum_{i=1}^{t} \Phi_x^{t,t-i} \tilde{w}_{i-1}$. Then, the following constraints
\vspace{-0.2cm}
\begin{equation} \label{eq:tightened_constr}
\vspace{-0.2cm}
    \small a_j^{x\top} h_t + \sum_{i=1}^{t} \left\lVert a_j^{x\top} \Phi_x^{t,t-i} \right\rVert_1 \leq b_j^x, \quad j = 1,\dotsc,n_{\mathcal{X}}
\end{equation}
% comment: need to define \epsilon_x^t(i)
guarantee that all constraints in $\mathcal{X}$ are satisfied robustly. As discussed in Section~\ref{sec:problem_formulation}, the recursive feasibility of the linear PSF cannot be guaranteed without a robust forward invariant set. Therefore, in this work, we introduce soft constraints into \eqref{eq:tightened_constr},
\vspace{-0.2cm}
\begin{equation} 
\vspace{-0.1cm}
\label{eq:soft_state}
\small
\begin{aligned}
   a_j^{x\top} h_t + \sum_{i=1}^{t} \left\lVert a_j^{x\top} \Phi_x^{t,t-i} \right\rVert_1 \leq b_j^x + \epsilon_{x,j}^t,\quad&\epsilon_{x,j}^t \geq 0,\, t \in [T],\\
   & j = 1,\dotsc,n_{\mathcal{X}}.
\end{aligned}        
\end{equation}
In the cost function, a large penalty on $\lVert \epsilon_x^t \rVert_1$ is applied. In the case $\epsilon_{x,j}^t = 0$, we obtain robust constraint satisfaction guarantee for the constraint with parameters $(a_j^x, b_j^x)$. Similarly, the input constraints can be tightened as
\vspace{-0.1cm}
\begin{equation} \label{eq:soft_input}
\vspace{-0.1cm}
\small
\begin{aligned}
   a_k^{u\top} v_t + \sum_{i=1}^{t} \left\lVert a_k^{u\top} \Phi_u^{t,t-i} \right\rVert_1 \leq b_k^u + \epsilon_{u,k}^t,\quad&\epsilon_{u,k}^t \geq 0,\, t \in [T],\\
   & k = 1,\dotsc,n_{\mathcal{U}},
\end{aligned}        
\end{equation}
where $n_\mathcal{U}$ denotes the number of linear inequalities defining $\mathcal{U}$ and $(a_k^u, b_k^u)$ denotes the $k$-th set of linear constraint parameters in $\mathcal{U}$. The trust region constraints $(x_t, u_t) \in \mathcal{B}_\infty \left(\hat{z}_t, r_t\right)$ analogously can be tightened as
\vspace{-0.1cm}
\begin{equation} \label{eq:soft_trust_region}
\small
\begin{aligned}
   &a_j^\top h_t + \sum_{i=1}^{t} \left\lVert a_j^\top \Phi_x^{t,t-i} \right\rVert_1 \leq b_j + \sigma_{x,j}^t,\, \sigma_{x,j}^t \geq 0, \\ 
    &a_k^\top v_t + \sum_{i=1}^{t} \left\lVert a_k^\top \Phi_u^{t,t-i} \right\rVert_1 \leq b_k + \sigma_{u,k}^t,\, \sigma_{u,k}^t \geq 0, \\ 
   & j = 1,\dotsc,2 n_x,\, k = 1,\dotsc, 2n_u,\,  t \in [T-1],
\end{aligned}
\end{equation}
where $(a_j, b_j)$ and $(a_k, b_k)$ are defined similarly for the state and input-related constraint parameters in $\mathcal{B}_\infty \left(\hat{z}_t, r_t\right)$. 

Summarizing the results above, we propose a convex tightening of the linear PSF, which can be written as
\vspace{-0.2cm}
\begin{equation} \label{eq:convex_filter}
\begin{aligned}
	\underset{\substack{\Phix, \Phiu, \mathbf{\Psi}, \hh, \vv, \\ \{ \epsilon_x^t, \epsilon_u^t, \sigma_x^t, \sigma_u^t \} } }{\textrm{minimize}} & \quad \left\lVert u_0 - \pi(x(k)) \right\rVert_2^2 + M_\epsilon \sum_{t=0}^T \left(\left\lVert \epsilon_x^t \right\rVert_1 + \left\lVert \epsilon_u^t \right\rVert_1\right) \\
        & \quad + M_\sigma \sum_{t=0}^{T-1} \left(\left\lVert \sigma_x^t \right\rVert_1 + \left\lVert \sigma_u^t \right\rVert_1 \right)\\
	\textrm{subject to} & \quad \text{affine constraint~\eqref{eq:affine_scaled}},\\
        & \quad \text{over-approximation constraints~\eqref{eq:over_approx_constr}}, \\
        & \quad \text{soft state and input constraints~\eqref{eq:soft_state} and~\eqref{eq:soft_input}}, \\
        & \quad \text{soft trust region constraints~\eqref{eq:soft_trust_region}},\\
	& \quad x_0 = x(k),
\end{aligned}
\end{equation}
where $M_\epsilon, M_\sigma > 0$ are chosen as large numbers. When a polyhedral robust forward invariant set is used as the terminal constraint, we can tighten it similarly as~\eqref{eq:soft_state}. All the constraints in Problem~\eqref{eq:convex_filter} are linear, making \eqref{eq:convex_filter} a quadratic program. %The constants $M_\epsilon$ and $M_\sigma$ are chosen large, in order to drive the slack variables to zero. Furthermore, they can be used to prioritize the constraint satisfaction of either the trust region constraints or the safety constraints. 
With the use of soft constraints, \eqref{eq:convex_filter} is always feasible. If the slack variables $\{ \epsilon_x^t, \epsilon_u^t, \sigma_x^t, \sigma_u^t \}$ in the solution are zero, we obtain a certificate that the system is safe for the next $T$ steps under the state-feedback controller $\uu = \KK(\xx - \hh ) + \vv$ with $\KK = \Phiu \Phix^{-1}$.

\subsection{Trust region update}
\label{sec:trust_region_update}
Following the discussion on the trust regions in Section~\ref{sec:nnv_bounds}, we describe a method to update the trust regions online. Starting with the reference trajectory given by the primary policy $\pi(x)$, we propose to iteratively increase $r_t$ and update the reference trajectory by applying the policy $\uu = \KK (\xx - \hh) + \vv$, synthesized in Section \ref{sec:robust_mpc}. We then pick the reference trajectory that gives the smallest slack variables and apply the corresponding control inputs to the system. Our framework is summarized in Algorithm~\ref{alg:filter}.

\begin{algorithm}[htb]
\caption{Robust Linear MPC-based PSF}\label{alg:filter}
\textbf{Input:} Current state $x(k)$, horizon $T$, number of iterations $N$, initial reference trajectory $\hat{z}_{0:T}$, reference control input $\pi(x(k))$, initial trust region radius $r_0 >0$ \\
\textbf{Output:} Filtered control input $u_0^*$, safety certificate \texttt{safe\_cert}
\begin{algorithmic}[1]
\State \textit{At each time step} $k$,
\State \textbf{Initialize:} $r_t \leftarrow r_0,\;t \in [T-1], \{\KK^*, \hh^*, \vv^* \} \leftarrow \{\textbf{0}\}$, $\epsilon^* \leftarrow \infty$.
\For{$\ell = 1, \dots, N$}
\State Construct $\mathcal{B}_\infty(\hat{z}_t, r_t)$ and $\mathcal{P}_t$ with \eqref{eq:linear_bounds_direct} and \eqref{eq:linear_bounds_shift}
\State Solve \eqref{eq:convex_filter} with $x(k)$ to get $\{\KK, \hh, \vv \}$
\State Extract $\epsilon_{max}:=$ max. value of slack variables
\If{$\epsilon_{max} = 0$}
\State Set $\{\KK^*, \hh^*, \vv^* \} \leftarrow \{\KK, \hh, \vv \}$
\State Set \texttt{safe\_cert} $\leftarrow$ \texttt{True}
%\State Set $\ell \leftarrow N$ \textit{// early termination}
\Else
\State Update $\hat{z}_{0:T}$ with $\uu = \KK (\xx - \hh) + \vv$, as described  
\Statex $\quad\quad\;\;$ in \eqref{eq:ref_traj}
\State Update $r_t \leftarrow \beta r_t$ for $t \in [T-1]$ with $\beta > 1$.
\If{$\epsilon_{max} \leq \epsilon^*$}
\State Set $\epsilon^* \leftarrow \epsilon_{max}$
\State Set $\{\KK^*, \hh^*, \vv^* \} \leftarrow \{\KK, \hh, \vv \}$\State Set \texttt{safe\_cert} $\leftarrow$ \texttt{False}
\EndIf
\EndIf
\EndFor
\State Set $u_0^* \leftarrow v_0$, extracted from $\{\KK^*, \hh^*, \vv^* \}$ 
\end{algorithmic}
\end{algorithm}

% simulation 
\section{Numerical Example}
\label{sec:simulation}
To verify the efficacy of the proposed solution, we test it on a NN proxy of the nonlinear pendulum system~\footnote{ Our codes are publicly available at \url{https://github.com/ShaoruChen/NN-System-PSF}.}. The pendulum consists of the following dynamics \cite{brockman2016openai},
\begin{equation}\label{eq:pen_dynamics}
\ddot{\theta} = \frac{3g\sin(\theta)}{2l} + \frac{3\tau}{ml^2}\,,
\end{equation}
where $\theta$ is the angle between the pendulum and the vertical, $m$ and $l$ are the mass and length of the pendulum, $g$ is the gravitational force, and $\tau$ is the external torque acting on the pendulum. The state and control input are defined as $x := [\theta\; \dot{\theta}]^{\top} \in \mathbb{R}^2$ and $u := \tau \in \mathbb{R}$. To obtain the linear dynamics in \eqref{eq:dynamics}, the dynamics \eqref{eq:pen_dynamics} are linearized about the origin and discretized with a sampling time of 0.05s to obtain the following dynamics,
\begin{equation}\label{eq:pen_dynamics_linear}
	x_{t+1} = \begin{bmatrix} 1.0092 & 0.05015 \\ 0.369 & 1.0092 \end{bmatrix} x_t + \begin{bmatrix} 0.00125 \\ 0.05015 \end{bmatrix} u_t.
\end{equation}
\setlength{\textfloatsep}{0.1cm}

Next, we train a NN $f(x, u)$ to approximate the residual dynamics that are not captured by the linear dynamics in \eqref{eq:pen_dynamics_linear}. We first collect data by simulating the nonlinear dynamics in \eqref{eq:pen_dynamics} for a duration of 15s. The NN is then trained through a backpropagation procedure \cite{paszke2017automatic}, using the mean squared errors between the predicted and true states as the loss function. The NN consists of 3 hidden layers with 64 neurons in each layer and uses the rectified linear unit (ReLU) as the activation function. Additive noise with a maximum magnitude $\sigma_w$ of $\{0.05,0.1\}$ is injected into the states of the system. For the primary control policy, we consider an iterative linear quadratic regulator (iLQR) scheme \cite{li2004iterative} with the box-constrained heuristic \cite{tassa2014control}. This is implemented with the \textit{mpc.pytorch} library \cite{amos2018differentiable}. The box-constrained heuristic allows the system to adhere to the control constraints, but does not account for state constraints. 

Four test cases are considered in our experiments. In each of these cases, the system is required to track a pair of reference angles $\left(\theta_{r,1},\,\theta_{r,2}\right)$ sequentially, starting from an initial condition $x_0$ and across a duration of 2s. The initial conditions and reference angles are given in Table \ref{tab:ic_ref}.
\begin{table}
\vspace{0.2cm}
\centering
\caption{Initial conditions and angles for the test scenarios.
\label{tab:ic_ref}}
\vspace{-0.3cm}
\resizebox{0.9\columnwidth}{!}{
\begin{tabular}{*5c}
\toprule
Test Case & $x_0$ [deg; deg/s] & $\theta_{r,1}$ [deg] & $\theta_{r,2}$ [deg]\\
\midrule
1 & [57.3;\;-120.3] & 120 & -50\\
2 & [-85.9;\;-85.9] & -150 & 40\\
3 & [-85.9;\;-114.6] & -100 & -180\\
4 & [85.9;\;57.3] & 100 & 180\\
\bottomrule
\end{tabular}
}
\end{table}
We simulate each of these test cases under 4 control schemes - (i) a nominal iLQR framework, (ii) a soft-constrained iLQR framework (SC-iLQR), (iii) safe-filtered iLQR, where we apply the proposed safety filter to the nominal iLQR scheme, and (iv) safe-filtered SC-iLQR, where the safety filter is applied to SC-iLQR. For SC-iLQR, soft state constraints are incorporated into the cost function of the forward pass of the iLQR algorithm. Specifically, the function $\phi(x) =\text{max}\{0,x\}$ is applied onto each of the constraints, which increases the cost function proportionally whenever the constraints are violated. In safe-filtered iLQR and SC-iLQR, the initial reference trajectories $\hat{z}_{0:T}$ in Algorithm~\ref{alg:filter} were initialized by iLQR and SC-iLQR, respectively. 

\begin{figure}
    \center
    \includegraphics[scale=0.45, trim = -0.5cm 0.5cm 0cm -0.4cm]
    {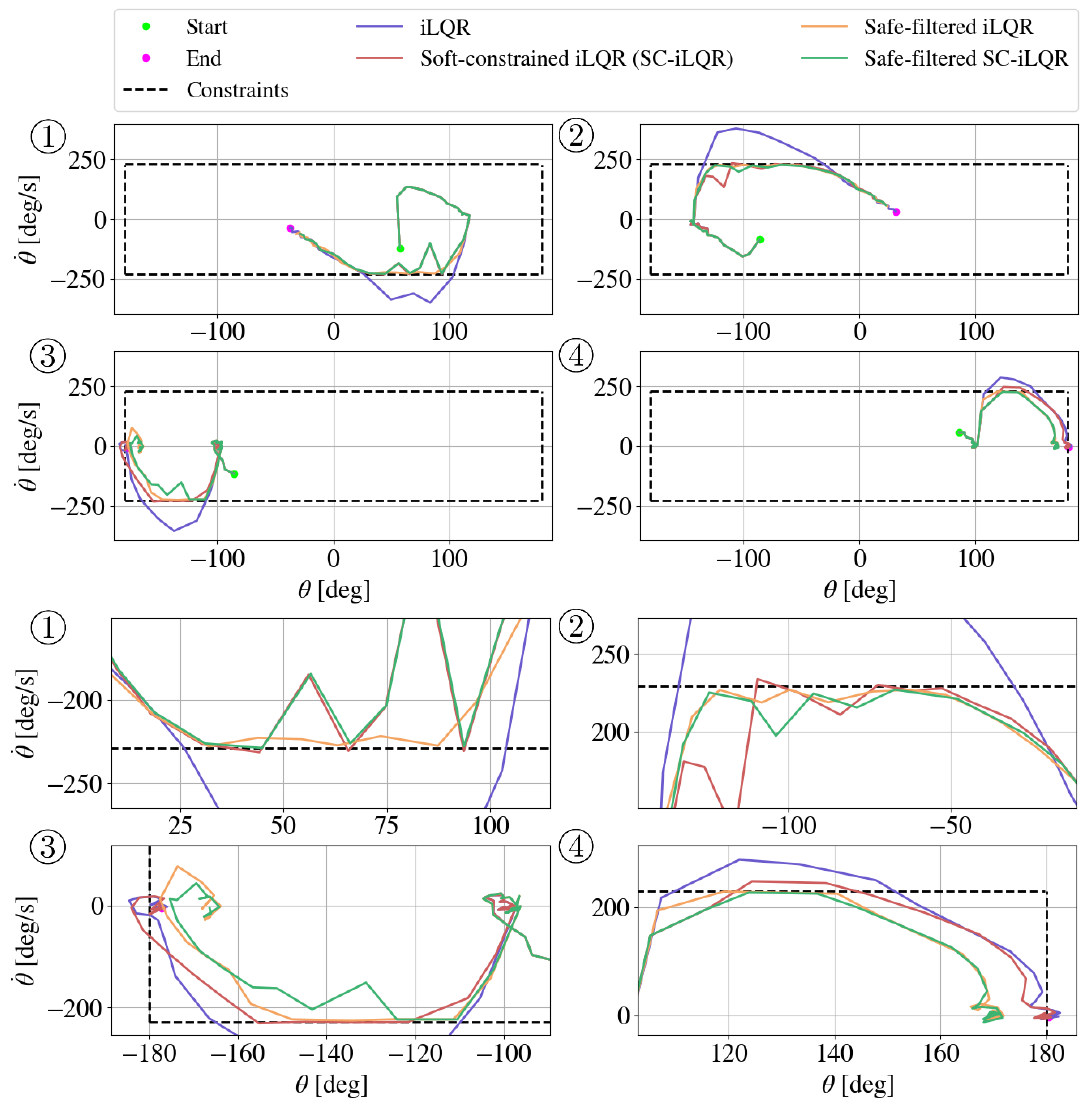}
    \caption{\textbf{Top panel}: Plots of the state trajectories under the 4 test cases. The initial and final states are marked with green and magenta circles. The state constraints are plotted with black dashed lines. \textbf{Bottom panel}: Zoomed-in plots of the state trajectories, around the boundaries where constraint violation potentially occurs.}
    \label{fig:comparison}
\end{figure}

\begin{figure}
    \center
    \includegraphics[scale=0.245, trim = -0.5cm 1.5cm 0cm -0.0cm]
    {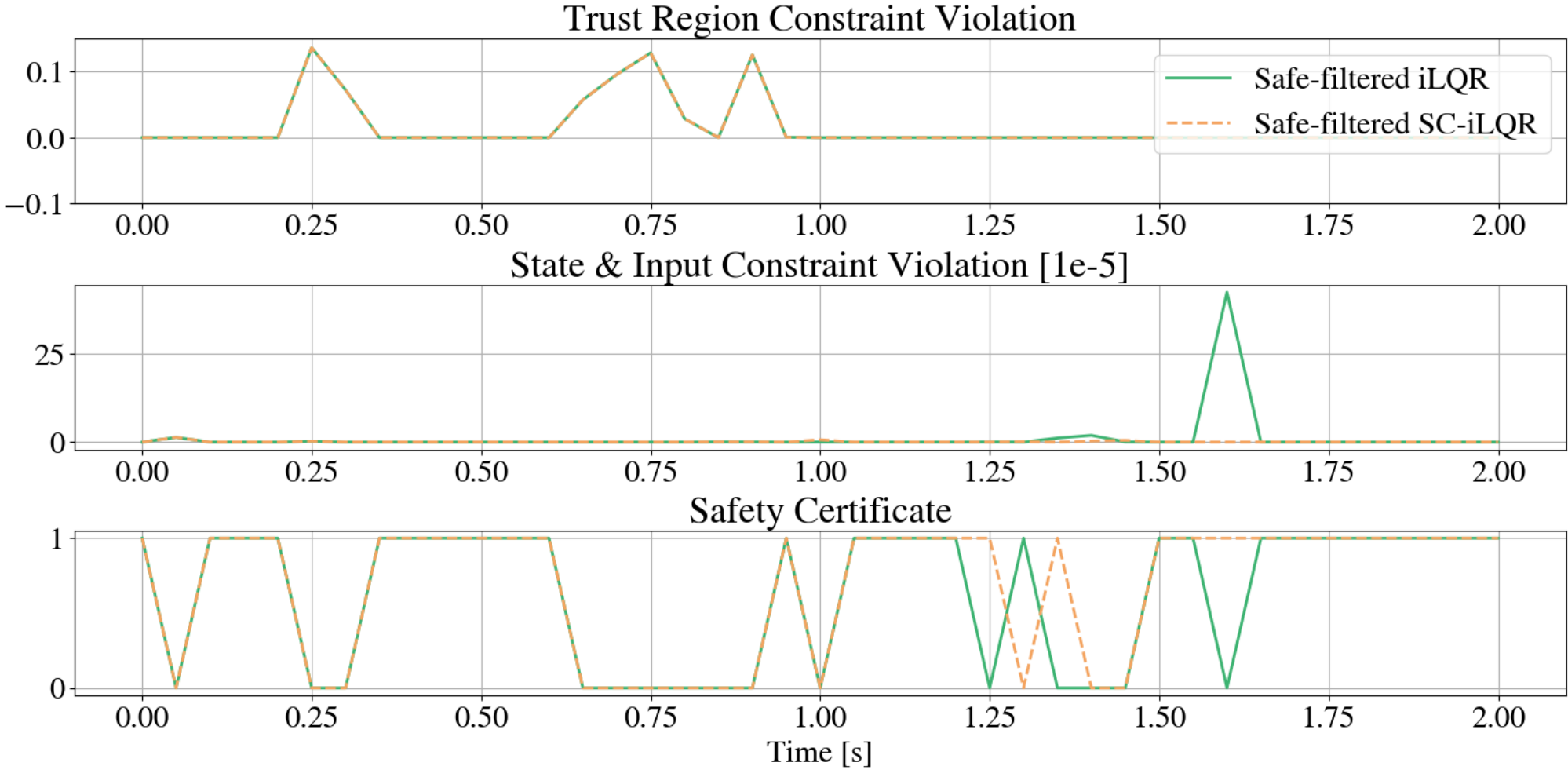}
    \caption{\textbf{Top two panels}: Time histories of the slack variables for the trust region, state and input constraints. \textbf{Bottom panel}: Time histories of the safety certificate attained. The values of 1 and 0 denote \texttt{True} and \texttt{False} respectively.}
    \label{fig:constraints}
\end{figure}

The state trajectories under these control schemes are plotted in Fig. \ref{fig:comparison} and the percentages of constraint violations are tabulated in Table \ref{tab:const_violate}. These percentages are computed by taking the ratio between the number of points in which the states violate the constraints against the total number of points in the state trajectory. Since the nominal iLQR method does not account for state constraints, it results in the largest percentage of constraint violation. While the soft-constrained iLQR reduces the level of constraint violation, there are a number of instances where the constraints are violated, as depicted in Fig. \ref{fig:comparison}. On the other hand, as shown in Table \ref{tab:const_violate}, through the application of the safety filter to iLQR and SC-iLQR, no constraint violations are observed for the test cases. To illustrate the safety certificate obtained, we plot the slack variables that characterize the trust region and state and input constraints, together with the safety certificate for the third test case, under a maximum noise level $\sigma_w=0.05$, in Fig. \ref{fig:constraints}. {The combination of Table~\ref{tab:const_violate} and Fig.~\ref{fig:constraints} indicates that given the formulation in~\eqref{eq:convex_filter}, the PSF may not give a numerical safety certificate even when the state trajectories are safe. Meanwhile, this proposed PSF can effectively encourage the system to behave safely by minimizing the conservative upper bounds on the constraint violation. }

\begin{figure}
    \center
    \includegraphics[scale=0.25, trim = 1cm 1cm 0cm 0cm]
    {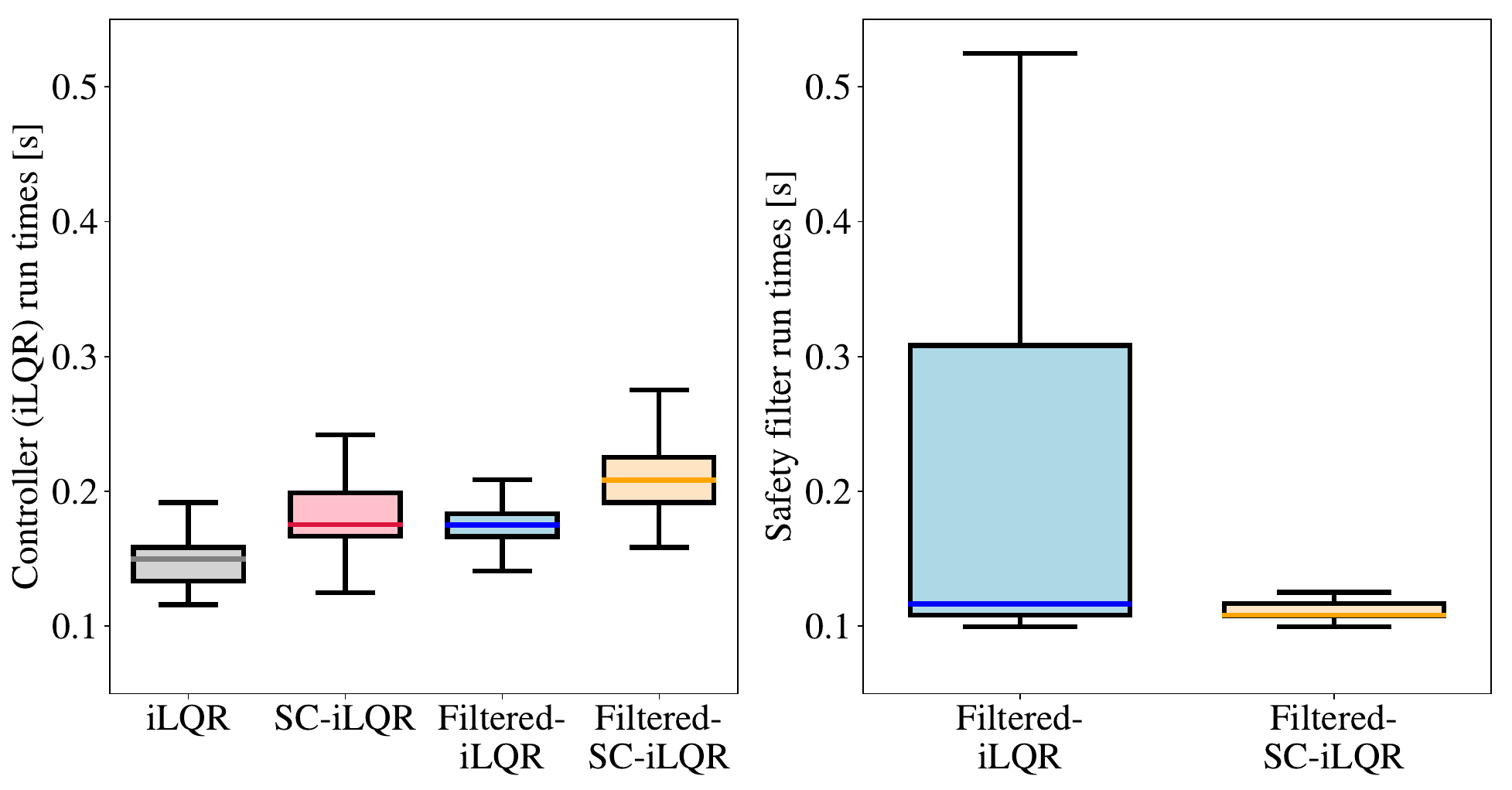}
    \caption{Statistics of the run times of the controller and the safety filter, under the four test methods.}
    \label{fig:compute_time}
\end{figure}

The statistics of the computational times of the control scheme, in this case the iLQR scheme, and the safety filter are depicted in Fig. \ref{fig:compute_time}. While introducing soft constraints increases the run times of the iLQR scheme, it has a significant effect on the run times of the safety filter, as observed in the right panel of Fig. \ref{fig:compute_time}. With the soft constraints added to the iLQR scheme, the state trajectories are closer to the boundaries of the constraint sets, without the activation of the safety filter. This allows the filter to find a solution that satisfies the constraints under a smaller number of iterations, which reduces computational time.

\begin{table}
\vspace{0.2cm}
\centering
\caption{Percentage of state constraint violations.
\label{tab:const_violate}}
\resizebox{\columnwidth}{!}{
\begin{tabular}{*9c}
\toprule
Method & \multicolumn{4}{c}{Maximum noise level, $\sigma_w=0.05$} \\
{} & \multicolumn{1}{c}{Case 1 [\%]} & \multicolumn{1}{c}{Case 2 [\%]} & \multicolumn{1}{c}{Case 3 [\%]} & \multicolumn{1}{c}{Case 4 [\%]}\\
\midrule
iLQR & 12.20 & 14.63 & 14.63 &  21.95\\
SC-iLQR & 7.32 & 4.88 & 12.20 & 12.20\\
Safe-filtered iLQR & \textbf{0.0} & \textbf{0.0} & \textbf{0.0} & \textbf{0.0} \\
Safe-filtered SC-iLQR & \textbf{0.0} & \textbf{0.0} & \textbf{0.0} & \textbf{0.0} \\ 
\midrule
{} & \multicolumn{4}{c}{Maximum noise level, $\sigma_w=0.1$}\\
{} & \multicolumn{1}{c}{Case 1 [\%]} & \multicolumn{1}{c}{Case 2 [\%]} & \multicolumn{1}{c}{Case 3 [\%]} & \multicolumn{1}{c}{Case 4 [\%]}\\
\midrule
iLQR & 14.63 & 14.63 & 17.07 &  26.83\\
SC-iLQR & 4.87 & 2.44 & 7.32 & 9.75\\
Safe-filtered iLQR & \textbf{0.0} & \textbf{0.0} & \textbf{0.0} & \textbf{0.0} \\
Safe-filtered SC-iLQR & \textbf{0.0} & \textbf{0.0} & \textbf{0.0} & \textbf{0.0} \\ 
\midrule
\bottomrule
\end{tabular}
}
\end{table}

% \begin{table}
% \vspace{0.2cm}
% \centering
% \caption{Computational time in seconds.
% \label{tab:const_violate}}
% \resizebox{\columnwidth}{!}{
% \begin{tabular}{*9c}
% \toprule
% Method & \multicolumn{1}{c}{T_{\text{control,avg}}} & \multicolumn{1}{c}{T_{\text{control,median}}} & \multicolumn{1}{c}{T_{\text{filter,avg}}} & \multicolumn{1}{c}{T_{\text{filter,median}}}\\
% \midrule
% iLQR & 0.145 & 0.150 & - &  -\\
% SC-iLQR & 0.182 & 0.175 & - & -\\
% Safe-filtered iLQR & 0.174 & 0.175 & 0.250 & 0.116 \\
% Safe-filtered SC-iLQR & 0.212 & 0.208 & 0.136 & 0.108 \\ 
% \bottomrule
% \end{tabular}
% }
% \end{table}
% conclusion
\section{Conclusion}
\label{sec:conclusion}
We propose a convex optimization-based predictive safety filter for uncertain NN dynamical systems with the inclusion of additive disturbances. By utilizing tools from NN verification and robust linear MPC, our method requires solving a soft-constrained convex program online whose complexity is independent of the NN size. With our framework, formal safety guarantees, together with a robust state-feedback controller, are attained when the slack variables in the solution are all zero. 
% For future work, we plan to improve the implementation of the safety filter in terms of computational speed, to enable its potential for real-world applications.

\appendix
\subsection{Proof of Proposition~\ref{prop:sufficient}}
\label{app:proof}
Showing that $\mathcal{R}(\bfeta; \KK, \hh, \vv) \subseteq \mathcal{R}(\mathbf{\Psi} \tildeww)$ is equivalent to showing for every possible values of $\bfeta$, there exist $\tildeww^* \in \widetilde{\mathcal{W}}$ such that $\bfeta = \mathbf{\Psi} \tildeww^*$. Following this perspective, the sufficient conditions in Proposition~\ref{prop:sufficient} can be derived inductively. 

\paragraph{Initial case}
At $t = 0$, we have $x_0 = h_0, u_0 = v_0$, and 
\begin{equation} \label{eq:bounds_on_Delta_0}
\begin{aligned}
\Delta_0(x_0, u_0) &\geq \begin{bmatrix}\ubar{D}_0^x & \ubar{D}_0^u \end{bmatrix} y_0 + \underline{d}_0, \\
\Delta_0(x_0, u_0) &\leq \begin{bmatrix} \obar{D}_0^x  & \obar{D}_0^u \end{bmatrix} y_0 + \overline{d}_0,
\end{aligned}
\end{equation}
with $y_0 = [h_0^\top \ v_0^\top]^\top$. Note that $\eta_0 = \Delta_0(x_0, u_0) + w_0$. For $\eta_0 = \Psi^{1,0} \tilde{w}_0 = \diag(\psi_0)\tilde{w}_0$ to have a solution $\tilde{w}_0$ satisfying $\lVert \tilde{w}_0 \rVert_\infty \leq 1$ for all possible realization of the lumped uncertainty $\eta_0$, it is both sufficient and necessary to have
\begin{equation} \label{eq:cond_0}
    \lVert \diag(\psi_0)^{-1} \eta_0 \lVert_\infty \leq 1 \Leftrightarrow  \left \{ \begin{array}{c}
         e_i^\top \eta_0 \leq \psi_{0,i}, i \in [n_x], \\
         -e_i^\top \eta_0 \leq \psi_{0,i}, i \in [n_x],
    \end{array} \right.
\end{equation}
hold robustly for $\eta_0$ where $e_i$ denotes the $i$-th standard basis. Using the bounds in~\eqref{eq:bounds_on_Delta_0} and the fact that $\lVert w_0 \rVert_\infty \leq \sigma_w$, we have constraints~\eqref{eq:over_approx_0} and~\eqref{eq:over_approx_1} guarantee the robust feasibility of~\eqref{eq:cond_0} for all possible values of $\eta_0$. For any realization of $\eta_0$, we denote the corresponding solution as $\tilde{w}_0^*$ such that $\eta_0 = \diag(\psi_0) \tilde{w}_0^*$.

\paragraph{Induction step}
For a given controller $\{\KK, \hh, \vv\}$, consider an arbitrary realization of the lumped uncertainty $\eta_{0:T}$. At time $t \geq 1$, let $\bfeta_{0:t} := [\zero^\top \ \eta_0 \ \cdots \ \eta_{t-1}^\top]^\top$ denote the truncation of $\bfeta$ consisting of the first $t+1$ components of $\bfeta$, and $\mathbf{\Psi}_{0:t} \in \mathcal{L}_{TV}^{t}$ denote the truncation of $\mathbf{\Psi}$ up to the $t+1$-th row and column. The truncated vectors $\mathbf{x}_{e, 0:t}, \mathbf{u}_{e, 0:t}$ and matrices $\mathbf{A}_{0:t}, \mathbf{B}_{0:t}$ are defined similarly. Assume there exist $\tildeww^*_{0:t} = [\zero^\top \ \tilde{w}_0^{* \top} \ \cdots \ \tilde{w}_{t-1}^{* \top}]^\top$ such that $\bfeta_{0:t} = \mathbf{\Psi}_{0:t} \tildeww^*_{0:t}$. Next, we will show that under constraint~\eqref{eq:over_approx_constr} there exist $\tilde{w}^*_t$ with $\lVert \tilde{w}^*_t \rVert_\infty \leq 1$ such that 
\begin{equation} \label{eq:robust_equation}
\begin{aligned}
    \eta_t &= \sum_{i=1}^{t+1} \Psi^{t+1,t+1-i} \tilde{w}_{i-1}^* \\
    &= \sum_{i=1}^{t} \Psi^{t+1,t+1-i} \tilde{w}_{i-1}^* + \diag(\psi_t)\tilde{w}_t^*
\end{aligned}
\end{equation}
holds.

Since $\bfeta_{0:t} = \mathbf{\Psi}_{0:t} \tildeww^*_{0:t}$, the error dynamics~\eqref{eq:error_dynamics} up to time $t$ can be written as 
\begin{equation} \label{eq:error_w_star}
       \mathbf{x}_{e, 0:t} = Z (\mathbf{A}_{0:t} \mathbf{x}_{e, 0:t} + \mathbf{B}_{0:t} ) \mathbf{u}_{e, 0:t} + \mathbf{\Psi}_{0:t} \tildeww^*_{0:t}.
\end{equation}
According to~\cite[Corollary 1]{chen2022robust}, the affine constraint~\eqref{eq:affine_scaled} parameterizes all closed-loop system responses $\mathbf{x}_{e, 0:t} = \mathbf{\Phi}_{x, 0:t} \tildeww^*_{0:t}$, $\mathbf{u}_{e, 0:t} = \mathbf{\Phi}_{u,0:t} \tildeww^*_{0:t}$ under the controller $\mathbf{u}_{e, 0:t} = \KK_{0:t} \mathbf{x}_{e, 0:t}$ where $\mathbf{\Phi}_{x, 0:t}, \mathbf{\Phi}_{u, 0:t}, \KK_{0:t} \in \mathcal{L}_{TV}^{t}$ are the corresponding truncation of $\Phix, \Phiu, \KK$, respectively. Define $y_t := [h_t^{\top}\,v_t^{\top}]^{\top}$ and $\Phi^{t,t-i} := \begin{bmatrix} \Phi_x^{t,t-i \top} & \Phi_u^{t,t-i \top}\end{bmatrix}^{\top}$. It follows from~\eqref{eq:error_w_star} and the fact $\mathbf{x}_{e, 0:t} = \mathbf{\Phi}_{x, 0:t} \tildeww^*_{0:t}$, $\mathbf{u}_{e, 0:t} = \mathbf{\Phi}_{u,0:t} \tildeww^*_{0:t}$ that the state $x_t$ and control input $u_t$ at time $t$ under the policy $\{\KK, \hh, \vv\}$ are given by 
\begin{equation} \label{eq:t_response}
\begin{bmatrix}
    x_t \\ u_t
\end{bmatrix} =
z_t + \sum_{i=1}^t \Phi^{t,t-i} \tilde{w}_{i-1}^*.
\end{equation}
Correspondingly, the uncertainty $\Delta_t(x_t, u_t)$ at time $t$ are bounded by 
\begin{equation} \label{eq:eta_bound}
    \begin{aligned}
        \Delta_t(x_t, u_t) &\geq 
        \begin{bmatrix}\ubar{D}_t^x & \ubar{D}_t^u \end{bmatrix} z_t + \underline{d}_t \\
        &= \begin{bmatrix} \ubar{D}_t^x & \ubar{D}_t^u \end{bmatrix} \Big(y_t + \sum_{i=1}^t \Phi^{t,t-i} \tilde{w}_{i-1}^* \Big) + \underline{d}_t, \\
        \Delta_t(x_t, u_t) &\leq \begin{bmatrix} \obar{D}_t^x  & \obar{D}_t^u \end{bmatrix} z_t + \overline{d}_t \\
        & = \begin{bmatrix} \obar{D}_t^x & \obar{D}_t^u \end{bmatrix} \Big(y_t + \sum_{i=1}^t \Phi^{t,t-i} \tilde{w}_{i-1}^* \Big) + \overline{d}_t,
    \end{aligned}
\end{equation}
which is achieved by plugging~\eqref{eq:t_response} into the definition of $\mathcal{P}_t$. Recall that $\eta_t = \Delta_t(x_t, u_t) + w_t$. For
\begin{equation} \label{eq:eta_t_equation}
      \eta_t = \sum_{i=1}^{t} \Psi^{t+1,t+1-i} \tilde{w}_{i-1}^* + \diag(\psi_t)\tilde{w}_t
\end{equation}
to have a solution $\tilde{w}_t$ such that $\lVert \tilde{w}_t \rVert_\infty \leq 1$, it is equivalent to require
\begin{equation}\label{eq:robust_inequality}
\begin{aligned}
    e_i^\top \Big(\eta_t - \sum_{i=1}^{t} \Psi^{t+1,t+1-i} \tilde{w}_{i-1}^*\Big) &\leq \psi_{t,i}, i \in [n_x],\\
    -e_i^\top \Big(\eta_t - \sum_{i=1}^{t} \Psi^{t+1,t+1-i} \tilde{w}_{i-1}^*\Big) &\leq \psi_{t,i}, i \in [n_x],
\end{aligned}
\end{equation}
hold. Since $\Delta_t(x_t, u_t)$ is bounded by~\eqref{eq:eta_bound}, $\lVert \tilde{w}_i^* \rVert_\infty \leq 1$ for $i \in [t-1]$ and $\lVert w_t \rVert_\infty \leq \sigma_w$, we have constraints~\eqref{eq:over_approx_2} and~\eqref{eq:over_approx_3} are sufficient to guarantee the inequalities~\eqref{eq:robust_inequality} hold. Then, we denote the solution of~\eqref{eq:eta_t_equation} as $\tilde{w}_t^*$ for the given realization of $\eta_t$. We repeat this process until $t=T$ and in this way construct the virtual disturbance $\tildeww^*$ such that $\bfeta = \mathbf{\Psi} \tildeww^*$. Since the realization of $\bfeta$ is chosen arbitrarily, we prove Proposition~\ref{prop:sufficient}. 
\vspace{-0.2cm}

\bibliographystyle{ieeetr}
\bibliography{refs}

\end{document}